\documentclass[aps,prd,twocolumn,superscriptaddress,aas_macros,nofootinbib]{revtex4-1}

\usepackage{graphicx}
\usepackage{subfigure}
\usepackage{amsmath} 

\usepackage[colorlinks=true,citecolor=blue,urlcolor=magenta,breaklinks]{hyperref}
\usepackage{natbib}
\usepackage[T1]{fontenc}
\usepackage{times}
\begin{document}
\title[Primordial Rotating Black Holes]{ Primordial Rotating Black Holes}


%
%
%
%
%
%
%
\def\simpropto{\lower.2ex\hbox{$\; \buildrel \propto \over \sim \;$}}
\def\ltsim{\lower.5ex\hbox{$\; \buildrel < \over \sim \;$}}
\def\gtsim{\lower.5ex\hbox{$\; \buildrel > \over \sim \;$}}

\author{J.A. de Freitas Pacheco}
\email[]{
jose.pacheco@oca.eu
}

\affiliation{Observatoire de la C\^ote d'Azur - Laboratoire Lagrange\\
06304 Nice Cedex - France}
 \author{Joseph Silk}
\email[]{silk@astro.ox.ac.uk}
\affiliation{Institut d'Astrophysique de Paris, UMR 7095 CNRS, 
Sorbonne Universit\'{e}, \\ 98 bis Boulevard Arago, Paris 75014, France}
\affiliation{Beecroft Institute of Particle Astrophysics and Cosmology, 1 Keble Road, University of Oxford, Oxford OX1 3RH, UK}
 \affiliation{Department of Physics and Astronomy, 3701 San Martin Drive, The Johns Hopkins University, Baltimore MD 21218, USA}

%

\begin{abstract}

Primordial  black holes formed in an early post-inflation matter-dominated epoch during preheating provide a novel pathway for a source of the dark matter that utilizes known physics in combination with plausible  speculations about the role of quantum gravity. Two cases are considered here: survival of Planck-scale relics and an early universe accretion scenario for formation of primordial black holes of  asteroid-scale masses.

\vspace{0.5cm}

\end{abstract}
\maketitle

\section{Introduction}

Standard Model relics do not have sufficient abundance to explain
the existence of dark matter in the universe, apart from  (sterile) neutrinos which erase small-scale fluctuations
via their free-streaming. To avoid excessive power suppression,  investigations have 
generally focused on particles generated via extensions of the Standard Model and, in particular, from
minimal supersymmetric theories. So far, no signal of supersymmetry has
been found in experiments performed with the Large Hadron Collider \cite{arcadi}
and possible results from direct 
search experiments are unconfirmed \cite{bernabei, aprile}.

Negative results have also been
obtained via indirect searches related to the detection of high energy neutrinos,  leptons, hadrons. and photons  that could have
been produced by the annihilation of dark matter particles  predicted in diverse environments \cite{srednicki}, including 
the sun \cite{choi},  the solar neighborhood \cite{cirelli},  the Galactic Center \cite{leane}
or  in nearby dwarf galaxies  \cite{abdalla,hoof}.

Confronted with these difficulties, a revival of the idea that primordial black holes (PBHs) could be dark
matter particles  has flourished in the recent literature. Although early investigations on the formation of PBHs have 
focused on the radiation-dominated phase just after reheating, some studies  have  raised the possibility that  the
scalar fields that drive inflation may be affected by gravitational instabilities (similar to the Jeans instability) that may lead to
black hole formation \cite{khlopov1,khlopov2,malomed} at the end of inflation when
the fields (or field) driving the expansion oscillate(s) around a local minimum of the potential before 
decaying into matter/radiation. During this short oscillatory phase, the energy density varies approximately as 
$\rho_{\phi} \propto a^{-3}$, i.e., it behaves like a "dust'' fluid. This short matter-dominated  era is advantageous for black hole formation
since pressure gradients are practically absent in density fluctuations, contrary to  the situation in  the post-reheating 
phase. The formation of these PBHs depends not only on the amplitude of the perturbations but also on the form of the field potential. 
Moreover, as discussed below,  a broad mass spectrum of PBHs is produced that results in substantial growth of the most massive PBHs by accretion of small PBHs before they have time to Hawking evaporate.

Two principal scenarios have been envisaged: in the first, only one field is responsible for driving the inflation, while in the second, an 
additional ''spectator'' field (or several such fields) may be also present. In the former scenario, the scalar field can fragment into lumps via gravitational
instability and form black holes under certain conditions \cite{cotner}. In the second picture, one or more spectator fields are present.
These fields are characterized by the fact that their energy densities are never dominant in comparison with that of the main field. However, their 
curvature power spectra dominate at small scales, favoring black hole formation \cite{tenkanen}. PBHs originating from these mechanisms
have, in general, masses larger than $10^{15}~\rm g$ and hence they survive the Hawking evaporation process. 

Here we are concerned with smaller mass PBHs  that would evaporate on short timescales, potentially leaving  
a Planck-sized remnant. The possible existence of these relics and their possible stability are the main purpose of the present investigation.
It is worth mentioning that PBHs of small masses  ($M \sim 10^{6}~g$) are expected to be formed in a hybrid inflationary scenario 
as discussed by \cite{garcia,chen}. Also in a single
field scenario, small mass  black holes  form via a resonant instability that develops during the oscillatory phase for
wavelengths that exit the Hubble radius near the end of inflation \cite{martin}. This mechanism is able to produce PBHs with masses 
around $\sim 10^{3}~\rm g$.  

We envisage two alternative scenarios for these PBHs. They 
may  evaporate but leave stable  Planck mass remnants, 
accounting  for a substantial part of dark matter \cite{martin}. Alternatively, if the mass range  is broad, accretion of many low mass  PBHs by the most massive PBHs will produce long-lived PBHs that can equally account for most of the dark matter by populating the 
only available PBH window, in the mass range $10^{17}-10^{22} $ g.

Here we consider a scenario in which PBHs  formed
by the aforementioned mechanisms, leaving extremal rotating remnants. Our motivation is based on recent 
investigations suggesting that PBHs formed in such a particular phase of the universe might have acquired 
significant angular momentum \cite{harada,kuhnel}. 

Surviving  Planck-scale objects can be described by a particle-like formalism \cite{nicolini1,nicolini2} that permits the existence of an extremal configuration associated with the ground state.  The PBHs formed  by accretion must lie above the accretion limit imposed by the diffuse gamma ray background.

Our paper is organized as follows: Planck-sized relics and their stability are discussed in Section 2 and
the particle-like formalism describing the black hole will be presented in Section 3. It will be shown that
the resulting horizon area is equally spaced in agreement with the early findings by
Bekenstein \cite{bekenstein} and that the Bohr correspondence principle is verified or, in other words, General 
Relativity is recovered at large quantum numbers.  A following section describes the PBH accretion scenario. Our  main conclusions are
given in Section 5. 
\section{Planck remnants}

The Hawking evaporation process has always been an obstacle for explaining the survival of PBHs with 
masses smaller than $\sim 10^{15}~g$. To avoid such a difficulty, MacGibbon \cite{macgibbon} postulated that Planck-sized black holes could be surviving 
remnants of such a process and could eventually constitute a substantial fraction of the dark matter. The existence 
of relics of the Hawking process is still a matter of debate since it  is related not only to the possibility of having Planck-scale black holes as dark matter but also to the disappearance of the singularity \cite{sabine,bonanno}. 

A possible resolution derived from string theory leads to a generalization of the 
Heisenberg uncertainty principle (or simply GUP). This  suggests that  Hawking evaporation ends when the black hole mass attains a 
minimal value of the order of the Planck mass \cite{chen,adler,faizal}. Such remnants are expected to not emit radiation/particles
despite  having an extremely high horizon temperature.
Black holes with an inner Cauchy horizon could be a possible resolution of  this problem, especially since in the extremal case, i.e., when the two
horizons coincide, the associated temperature is zero and Hawking emission is suppressed. 

In this situation, extremal
regular black holes described either by the Bardeen or the Hayward metrics \cite{bardeen,hayward} are able to produce Planck-sized
objects if their Riemann curvature invariant is comparable to that derived in terms of the volume operator in Loop Quantum Gravity
\cite{pacheco2,modesto}. A similar approach was adopted by Chamseddine et al. \cite{chamseddine} who
assumed that General Relativity is modified at curvatures near the Planck scale. As a result, they have obtained stable (zero Hawking
temperature) remnants with masses determined by the inverse limiting curvature. 

We note that 
in loop gravity the Hawking evaporation time scales as $m_{bh}^2,$ as opposed to $m_{bh}^3$ in  Einstein gravity. This potentially affects the mass range of PBHs that survive evaporation.

Extremal singular charged or rotating black holes described respectively by the Reissner-Nordstrom or the Kerr metrics have been also 
considered in the literature. In the former case,  debate about discharge by spontaneous pair production remains open. This 
process affects large black holes ($M \sim 10^{17}~g$) while other processes also affect  the charge evolution of smaller ones as 
emphasized by \cite{gibbons,sorkin}. The main difficulty in describing the discharge mechanism is to find an adequate 
transition from a space-time with a Cauchy horizon to another in which the inner horizon is absent \cite{sorkin2,herman}. 
More recently, Lehmann et al. \cite{lehmann}, searching for detectable effects of primordial Planck mass black holes, explored the idea 
that a fraction of them could be charged, regardless of their origin and stability.

\section{Quantum black holes}

A quantum description of a black hole  necessarily requires a complete quantum theory of gravitation, still non-existent. In the 
meantime, it is possible to adopt an approach in which either the mass-energy spectrum or the horizon surface of the black hole 
may assume discrete values. In fact,
in a seminal paper, Bekenstein \cite{bekenstein} considered that the horizon area of non-extremal black holes behaves as a classical adiabatic invariant, corresponding to a quantum operator with a discrete spectrum, according to the Ehrenfest principle. His point of departure was the
squared-irreducible mass relation for a Kerr-Newman black hole with which quantum mechanical operators are associated. Another point of view was
developed by Carr \cite{carr}, who assumed a smooth transition between the Schwarzschild and Compton scales associated with a black hole of
mass $M$. To some extent , these approaches suggest that BHs can be assimilated into elementary particles with Planck masses. Although these ideas can be pushed
a little further, one would expect that any formal quantum particle description of a black hole should permit the recovery
of General Relativity at larger quantum numbers. More recently, a particle-like black hole formalism was adopted \cite{nicolini2} to describe
a Reissner-Nordstrom (RN) black hole. This approach permits
the existence of extremal configurations that correspond to a minimal mass below which no horizon forms. The Hawking temperature
in this extremal state is zero, and objects deviating from extremal configurations radiate  well-defined quanta 
of energy in the form of  radiation and particles. Thus, one should expect that a quantum rotating black hole in an excited state
will also spontaneously decay into lower spin states since the asymmetric emission of neutrino quanta will carry away angular momentum
according to Leahy \& Unruh \cite{unruh}.

Here we will focus our interest in the Kerr space-time since it includes a Cauchy horizon permitting extremal solutions, stable from a
thermodynamic point of view. However, for the moment, the question concerning the existence of a ground-state solution that at the same time coincides with the extremal case remains open. It should be emphasized that black holes must also be considered as relativistic objects 
in the sense that they are subject to strong gravitational fields. Contrary to the approach by \cite{bekenstein}, who adopted the squared irreducible mass of a Kerr-Newman black hole as the departure point, here the relation 
defining  the inner ("$-$") and outer ("$+$") horizons of a Kerr black hole will be used, that is
\begin{equation}
\label{horizonkerr}
r_{\pm}(M,J) = \frac{GM}{c^2}\left[1 \pm \sqrt{1 - \left(\frac{Jc}{GM^2}\right)^2}\right]
\end{equation}
In the above equation, $M$ and $J$ are respectively the mass and the spin of the black hole and the other symbols have their usual meaning.
In a second step, it is assumed that the black hole spin is quantized by the usual rules, i.e., $J = m\hbar$, where $m = 0,1,2,...$.
In the particle-like description, the size of the black hole (or its outer horizon) should never be  smaller than the associated 
Compton wavelength. Hence, following \cite{nicolini2,pankovic}, one assumes that
\begin{equation}
\label{quantum1}
r_+(M,J) = n\frac{\hbar}{Mc}
\end{equation}
where $n = 1,2,3,...$ Notice that up to this point, nothing is imposed concerning the mass of the black hole, but as we 
shall see, the adopted quantum
conditions will lead naturally to Planckian values. Replacing the horizon radius given by eq.\ref{quantum1} into eq.\ref{horizonkerr}
permits us  to obtain an equation defining the possible values for the black hole masses, that is
\begin{equation}
\label{mass}
\frac{M}{M_P} = \sqrt{\frac{(n^2+m^2)}{2n}}
\end{equation}
where $M_P$ is the Planck mass. It is interesting to remark that the equation above, which gives discrete values for the black hole
mass, coincides with that derived by Bekenstein \cite{bekenstein} if the electric charge in his relation is put equal to zero. Now
insert  this result into eq.\ref{horizonkerr} and after some straightforward algebra, one obtains
\begin{equation}
\label{horizon1}
r_{\pm} = \ell_P\sqrt{\frac{(n^2+m^2)}{2n}}\left[1 \pm \sqrt{1-\frac{4m^2n^2}{(n^2+m^2)^2}}\right]
\end{equation}
where $\ell_P$ is the Planck length.

It can be shown trivially from eq.\ref{horizon1} that extremal black holes can be obtained if both quantum numbers are
equal, i.e.,  $n = m$. In this case, the horizon radius and the mass are respectively
\begin{equation}
r_H = \sqrt{n} \ell_P \,\,\,\, and \,\,\,\, M = \sqrt{n} M_P
\end{equation}

If such an extremal rotating black hole is in the ''ground'' state ($n = 1$), then the degenerate horizon $r_H$ and the mass
correspond precisely to the Planck length and mass with a spin equal to $\hbar$. Eq.\ref{horizon1} can be expanded and explicit
relations for the Cauchy and the event horizons  obtained, that is
\begin{equation}
\label{horizon2}
r_{-} = \sqrt{\frac{2m^4}{n(n^2+m^2)}}~\ell_P
\end{equation}
and
\begin{equation}
\label{horizon3}
r_{+} = \sqrt{\frac{2n^3}{(n^2+m^2)}}~\ell_P
\end{equation}

General Relativity imposes no restrictions on the mass spectrum of black holes, contrary to the discrete values resulting
from eq.\ref{mass}. If  Bohr's correspondence principle is applicable, one should expect to recover General Relativity
in the limit of large quantum numbers. Performing the ratio between eqs.\ref{horizon3} and \ref{mass}, one obtains after some
algebra
\begin{equation}
\frac{c^2r_{+}}{2GM} = \frac{n^2}{(n^2+m^2)}
\end{equation}
It can be easily verified that when  $n > m$ and $n >> 1$, the right side of the above equation  goes to unity and General Relativity
is recovered since the event horizon goes to $2GM/c^2$ as expected.

Another important aspect of the particle-like picture concerns the horizon area or the black hole entropy. Bekenstein, in his 
pioneering work \cite{bekenstein}, suggested that the black hole area should be discrete and equally spaced
with a spectrum $A_n = \alpha\ell_P^2n$ with $n = 1,2,...$ He estimated the 
proportionality constant as $\alpha = 8\pi$ by assuming that the squared irreducible mass of the black hole is the analogue of 
an adiabatic invariant action integral. Based on Bohr's correspondence principle, Hod \cite{hod} reached  a different
result for the proportionality constant, i.e., $\alpha = 4\ln 3$. This has been contested by other authors (see, for instance \cite{maggiore,ropotenko}), who confirmed the value $\alpha = 8\pi$ by independent analyses.

 It is interesting to verify if the present 
description of a quantum black hole is consistent with such a discretization of the black hole area and to ascertain the expected value of
the constant $\alpha$. For a Kerr black hole, the horizon area is given by
\begin{equation}
\label{area}
A = 4\pi\left[r_+^2 + \left(\frac{J}{Mc}\right)^2\right]
\end{equation}
Inserting into the above equation the horizon radius given by eq.\ref{horizon3}, the quantization condition for the spin and the black
hole mass given by eq.\ref{mass}, one  finally obtains
\begin{equation}
\label{area2}
A = 4\pi\ell_P^2\left[\frac{2n^3}{(n^2+m^2)} + \frac{2m^2n}{(n^2+m^2)}\right] = 8\pi\ell_P^2n
\end{equation}
Thus, the quantization conditions imposed in the present description of a quantum rotating black hole are consistent with
the discrete area spectrum first proposed by Bekenstein. It worth mentioning that Medved \cite{medved}, adopting a perturbative 
approach of the adiabatic invariant method, reached a similar result but included a fourth order spin correction term. Moreover,
for a Kerr black hole family of solutions, the Cauchy horizon is the future boundary of the domain of
dependence of the event horizon and the product of the areas of both horizons depends only on the squared angular momentum
\cite{marcus}. The area product $A_+A_-$ of both horizons can be simply written as
\begin{equation}
\label{product1}
A_+A_- = (8\pi)^2\left(\frac{GM}{c^2}\right)^2r_+r_-
\end{equation}
Inserting into this equation the expression for the mass and those for the inner and outer horizons, one obtains
\begin{equation}
\label{product2}
A_+A_- = (8\pi\ell^2)^2m^2 =\left(\frac{8\pi G}{c^3}\right)^2 J^2
\end{equation}
The last term on the right side of the above equation coincides with the relation given in \cite{marcus} when the electrical
charge is put equal to zero, being an additional test for the consistency of the particle-like model.

As we have seen previously, the "ground" state of the present quantum black hole model corresponds to $n = m = 1$, which also represents an extremal case. A further step is to examine the energy spectrum. For a Kerr black hole as well as for the Schwarzschild case, the 
total energy inside the horizon depends only on the black hole mass \cite{chamorro}. In this case, using the preceding results,
\begin{equation}
\label{energy}
E_{n,m} = M_Pc^2\sqrt{\frac{(n^2+m^2)}{2n}}
\end{equation}
Hence the ground state corresponds to an energy $E_{1,1} = M_Pc^2$, the Planck energy. A larger black hole (with $n \neq m$) 
will be in an excited state that can decay to a lower state by emitting an amount of energy $\Delta E = E_{n',m'}-E_{n,m}$. Such an energy quantum
$\Delta E$ does not correspond to a single particle (or photon) but should be imagined as a "bunch" of radiation/particles
with a well-defined total energy. If we appeal again to  Bohr's correspondence principle, one should expect that at large
quantum numbers, the equivalent transition frequency between two consecutive levels, i.e., $\omega = \Delta E/\hbar$ should 
approach the ''classical'' frequency representing a quasi-normal mode \cite{hod,maggiore}. In fact, when $n > m >> 1$, the energy gap between two consecutive levels is
\begin{equation}
\label{quantumenergy}
\Delta E_{n,m} = (E_{n+1,m+1}-E_{n,m}) \approx \frac{M_Pc^2}{\sqrt{2}}\left[\frac{1}{\sqrt{2n}} + \frac{m}{n^{3/2}}\right]
\end{equation}
In order to interpret the terms in the above equation, we recall that when $n > m >> 1$, the black hole mass is given by $M \approx M_P\sqrt{n/2}$. Thus, the first term on the right side of eq.\ref{quantumenergy} can be written $\hbar c^3/4GM$ or, equivalently, $2\pi~kT_H$ where $T_H$
is the Hawking temperature. The second term requires a little  extra work with the expression defining
the angular rotational velocity of a Kerr black hole (in geometric units), that is
\begin{equation}
\label{omega}
\Omega = \frac{4\pi a}{(r_+^2 + a^2)}
\end{equation}
where $a$ is the spin parameter. Performing the quantization as before and recovering the physical constants, the angular 
velocity can be recast as
\begin{equation}
\label{omega2}
\Omega_{n,m} = \frac{c}{2\ell_P}\frac{\sqrt{2}m}{\sqrt{n(n^2+m^2)}}
\end{equation}
For large values of the quantum numbers ($n > m$), one obtains  $\Omega_{n,m} \approx (c/\ell_P)(m/\sqrt{2n^3})$. Using these
results, eq.\ref{quantumenergy} can be rewritten as
\begin{equation}
\Delta E_{n,m} = \hbar\omega_{n,m} = 2\pi~kT_H + \hbar\Omega_{n,m}
\end{equation}
This expression is consistent with the limit at large quantum numbers for the real part of quasi-normal modes found by 
Hod \cite{hod} in the case of rotating black holes. On the other hand, for a black hole in a given state $(n, m)$ there is a limit
for the energy that can be extracted from the ergosphere \cite{ruffini}, which corresponds to the total utilization of the
rotational energy leaving the black hole in a state $(n, 0)$ or with an energy
\begin{equation}
E_{n,0} = M_Pc^2\sqrt{\frac{n}{2}}
\end{equation}
For consistency, the above energy must coincide with that corresponding to the irreducible mass  given by
\begin{equation}
E_{ir} = \frac{c^4}{G}\sqrt{\frac{A_+}{16\pi}} = M_pc^2\sqrt{\frac{n}{2}}
\end{equation}
where eq.\ref{area2} was used to replace the area of the horizon. Despite this
 mathematical consistency, one should 
expect that this result would be valid only for large black holes ($M > 10^{15}~g$) since for small ones 
the state $(n, 0)$  does not correspond to the ground state and the stability against the Hawking process disappears. Hence,
for small black holes, one should expect that the extraction of rotational energy stops when $m =1$. In this case, the
irreducible mass becomes
\begin{equation}
M_{irr}^2 = \frac{A_+c^4}{16\pi G^2} + \frac{4\pi\hbar^2}{A_+c^2}
\end{equation}
In the equation above, the second term on the right represents a "quantum correction" to the irreducible mass that permits
the black hole to decay to the ground state. Notice that for large black holes, such a correction term is completely negligible. 

\section{Accretion of PBHS by PBHs}
{
{Most discussions of PBH production argue that an extended mass function is fairly generic. Early production during an extended reheating phase boosts the PBH mass scale because of accretion by the rarer  more massive PBHs of the many smaller ones hovering near the ground state.  The generated mass function should peak near a mass limit determined by accretion over  the time available before reheating is completed. 
Using the spherical thin  shell approximation to accretion in an Einstein-de Sitter universe
(Bertschinger 1985), we estimate that the PBH masses  increase by  a factor $\sim (1+z_{inf})/(1+z_{rh}),
$
where $z_{inf}$ is the inflationary scale, estimated to be $\sim 3.10^{16}$ GeV for a weakly interacting and slowly  rolling scalar field that yields the observed low level of scalar  density fluctuations. Also,
 $z_{rh}$ is the reheat scale at a temperature estimated to be anywhere from $\sim 10^{16} \rm GeV$ down to the QCD scale 
 $\sim 100 \ \rm MeV$,    or even as low as  $\sim 10$ MeV  from BBN constraints. 
 
 Detection of primordial  B-mode CMB polarization offers our best way of determining the reheat temperature, but is constrained by restriction to the experimentally feasible regime of 
 $r \gtsim 0.0001$,  while  $r\simpropto T_{rh}^2 $    
allows a much larger  but unmeasurable range. 
Hence accretion allows PBHs produced during preheating to survive to the present epoch and  to populate the window accessible today by microlensing, provided they make up all or most of the dark matter of the universe.
There need be no adverse effects of early evaporation, eg during the epoch of nucleosynthesis, since the PBHs are mostly  evaporate later and are also very subdominant deep into the radiation era. Spectral $\mu$ distortions
of the CMB  produced via evaporation and energy injection into the thermal bath of the CMB  may provide a possible signature of their existence.
}}

\section{Concluding remarks}

Formation of PBHs in the very early universe, in particular during the short matter-dominated regime prevailing 
during the oscillatory phase of the inflaton field, has recently been the subject of extensive investigations.
Small mass black holes can be
formed in a hybrid inflation scenario first proposed by Linde \cite{linde} (see also \cite{chen}) as a consequence of a resonant 
instability affecting perturbations with wavelengths exiting the Hubble radius at the end of  inflation \cite{martin}. More
recently \cite{arya}, it was shown that PBHs with masses as small as $10^3~g$ can be formed in a warm inflation scenario since
the associated curvature power spectrum has a blue tilt that favors small-scale perturbations.

In these different scenarios, PBHs of small masses ($10^3~g$ of up to $10^{5}~g$) can be formed. Since these objects evaporate
in a short time-scale, the key question to be answered is whether or not a remnant will be left. Many authors have simply postulated the existence
of Planck-sized relics without any consideration about their nature or stability. If such relics are Planck-sized black holes,
their stability, at least from the thermodynamic point of view, is guaranteed if they are extremal black holes.
Regular or non-singular black holes described either by the Bardeen or the Hayward metrics are one possibility \cite{pacheco2}. However
these objects have a particular mass (or charge) distribution leading to a tenuous "outer" atmosphere, which is one reason to classify 
them more appropriately as "Planck stars" \cite{rovelli}. Moreover, instabilities appear in the core during the evaporation process that
complicate the analysis of the last stages of the process leading to remnant formation \cite{visser}. Extremal singular charged
or rotating black holes are an alternative possibility. While the former include, for instance, difficulties concerning the discharge by
the Schwinger mechanism, the latter  option is favored by recent studies on the spin distribution of PBHs, suggesting that a substantial fraction
has spin parameters near the maximum value \cite{harada}. Moreover, other studies \cite{arbey} indicate that rotating PBHs are not constrained by the
so-called Thorne limit \cite{thorne} and might have spin parameters in excess of $a \sim 0.998.$ 

If remnants of the evaporation process are Planck-sized black holes, one would expect that quantum effects become important. In fact,
a possible quantum approach was first proposed by \cite{bekenstein}, who considered that the horizon surface of a charged, rotating black hole
behaves as a classical adiabatic invariant having a discrete spectrum. Another quantum picture was adopted by \cite{nicolini2} to describe
charged black holes. In such a particle-like formalism it is possible to find extremal configurations associated with the ground state. Here
the particle-like formalism is adopted to describe rotating black holes. The ground state of such a particle-like Kerr black hole 
corresponds to a mass and a horizon having Planck sized values and a spin equal to $\hbar$. This state corresponds also to an 
extremal case. Excited states
satisfying the condition $n = m$ are also extremal but are expected to decay following excitation by mass accretion 
$n, m \rightarrow n', m$ with $n' > n$ 
and subsequent decay according to $n',m \rightarrow n'', m'$ with $n'', m' < n, m$. Similarly, the ground state population may oscillate due to these processes.    

Using the derived expressions for the horizon radius and mass, the resulting area is also quantized and equally spaced according 
to the original result first derived by Bekenstein \cite{bekenstein}. The energy spectrum is not equally spaced but satisfies  Bohr's correspondence principle, since the equivalent frequencies due to transitions between two consecutive levels at large quantum numbers 
approach the frequencies of quasi-normal modes. Moreover the product between the inner and the outer horizon surfaces is proportional
to the squared angular momentum showing that the quantum picture reproduces the main properties expected for macroscopic black holes.

{
Accretion during  reheating by the largest surviving  but rare PBHs of the far more numerous but small, down to Planck mass,  relic PBHs allows a substantial boost in PBH mass especially if reheating is late. This means that there are potentially observable signatures. These include  late evaporation into the diffuse gamma-ray and x-ray backgrounds \cite{auffinger}, cosmic ray signature in the (very) local interstellar medium \cite{boudaud} gravitational microlensing  of deep stellar fields \cite{subaru}, and spectral distortions of the CMB,  from evaporation in the radiation era ($\mu$-type) and in the matter era ($y$-type), as well as the intermediate hybrid ($\mu$-$y$) variety \cite{chluba}.
}

\end{document}